\documentclass[aps,prl,twocolumn,showpacs,a4paper]{revtex4}
\usepackage{amsmath,amsfonts}
\usepackage{graphicx}
\graphicspath{{figdistilled/}{./}}

\renewcommand{\vec}[1]{\ensuremath{\boldsymbol{#1}}}
\newcommand{\vhat}[1]{\ensuremath{\hat{\vec{#1}}}}
\newcommand{\topp}[1]{^{({#1})}}
\newcommand{\expval}[1]{\left\langle{#1}\right\rangle}

\newcommand{\per}{\ensuremath{\epsilon}}
\newcommand{\ez}{\ensuremath{F_z}}
\newcommand{\eznot}{\ensuremath{F_0}}
\newcommand{\gwidth}{\ensuremath{\Gamma}}
\newcommand{\dofstuff}{\ensuremath{d(\vhat{r},\vhat{n})}}
\newcommand{\pcan}{\ensuremath{P_{1,1}}}
\newcommand{\pfcan}{\ensuremath{\pf_{1,1}}}

\DeclareMathOperator{\bigo}{\mathcal{O}}

\DeclareMathOperator{\pf}{\tilde{P}} 

\newcommand{\concept}[1]{\emph{{#1}}}

\begin{document}

\title{%
The field inside a random distribution of parallel dipoles.
}
\date{\today}
\author{Janus H.~Wesenberg}
\author{Klaus M{\o}lmer}
\affiliation{%
  QUANTOP, Danish Research Foundation Center for Quantum Optics,
  Department of Physics and Astronomy, University of Aarhus, DK-8000
  {\AA}rhus C, Denmark
}
\begin{abstract}
  We determine the probability distribution for the field inside a
  random uniform distribution of electric or magnetic dipoles. 
  For parallel dipoles, simulations and an analytical derivation show
  that although the average contribution from any spherical shell
  around the probe position vanishes, the Levy stable distribution of
  the field is symmetric around a non-vanishing field amplitude.
  In addition we show how omission of contributions from a small
  volume around the probe leads to a field distribution with a
  vanishing mean, which, in the limit of vanishing excluded volume,
  converges to the shifted distribution.
\end{abstract}


\maketitle

\begin{figure}
  \centering
  \includegraphics{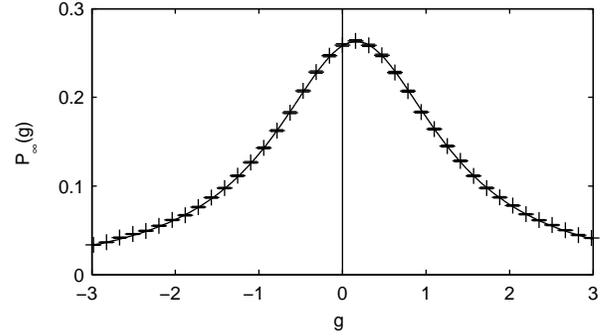}
  \caption{
    The probability distribution for $g=\ez/\eznot$, the scaled
    $z$-component of the field inside a random distribution of dipoles
    aligned along the $z$-axis.
    The symbols show the result of a numerical simulation (based on
    $10^6$ realizations of a system with $50.000$ individual dipoles,
    the uncertainty is represented by the line thickness),
    the solid curve is the exact Lorentzian solution for the
    probability distribution.
  }
  \label{fig:pardip2}
\end{figure}

The $z$-component of the field $\ez$ at the origin due to an electric or
magnetic dipole located at $\vec{r}$ is given by the expression
\begin{equation}
  \label{eq:ezfield}
  \ez=C  \frac{1}{r^3}
  \left(
    (\vhat{z}\cdot\vhat{n})
    -3(\vhat{r}\cdot\vhat{z})(\vhat{r}\cdot\vhat{n})
  \right),
\end{equation}
where $C=d/4\pi\varepsilon_0$ for an electric dipole
$\vec{d}=d\vhat{n}$ and $C=\mu_0m/4\pi$ for a magnetic dipole
$\vec{m}=m\vhat{n}$. $\vhat{z}$ and $\vhat{r}$ are unit vectors along
the z-axis and $\vec{r}$, respectively.
The field at a location within a random uniform distribution of many
dipoles is a superposition of terms like the one in Eq.(1).
The field component from a dipole parallel to the $z$-axis located at
a distance $r$ and at a direction $\theta$ with respect to the
$z$-axis is $C(1-3\cos^2\theta)/r^3$, and one sees that the average of
this expression over directions in space vanishes for all distances
$r$. It is hence surprising, that the field distribution in
Fig.~\ref{fig:pardip2}, obtained by numerical simulation, is
symmetrical around a non-vanishing value of the field. 
We shall prove analytically that the distribution is a shifted
Lorentzian, shown as the solid line in the figure, and that this is
the mathematical limit of distribution functions which all have
vanishing mean values but larger and larger variances.

The fields from electric and magnetic dipoles give rise to the most
important interactions of neutral matter, and they play significant
roles in atomic, molecular, and many-body physics. In the conclusion
we shall list topics in current quantum gas and quantum information
research where the present analysis may have important consequences.

A typical distance between dipoles with a given density $\rho$ is
$r_0=(3/4 \pi \rho)^{1/3}$, and a corresponding typical field strength
is $\eznot=C r_0^{-3}$. For notational convenience, we will rewrite
Eq.~\eqref{eq:ezfield} in terms of these typical values as
\begin{equation}
  \label{eq:gdef}
  g=\frac{\ez}{\eznot} =
  \left( \frac{r}{r_0}\right)^{-3} \dofstuff,
\end{equation}
where $\dofstuff$ is the geometrical factor of Eq.~\eqref{eq:ezfield}.

To derive the distribution function $P_N(g)$ for the field component
within a randomly distributed collection of $N$ dipoles we shall first
derive the distribution $P_{1,N}(g)=P(g|r<N^{1/3} r_0)$ for the
contribution from a single dipole within a sphere of radius $N^{1/3}
r_0$.
The combined field due to $N$ dipoles within the same sphere is
distributed according to the $N$-th order convolution product
\begin{equation}
  \label{eq:Pgn2}
  P_N(g)=\int \delta(\sum_{i=1}^N g_i-g)\ \prod_{j=1}^N
  P_{1,N}(g_j)\, dg_j.
\end{equation}

The probability distribution $P_{1,N}$ is calculated as
\begin{equation}
  \label{eq:probEz}
  P_{1,N}(g)=
  \expval{ \int_0^{N^{1/3} r_0} 
    \delta\left(
      g- \left(\frac{r}{r_0}\right)^{-3}\dofstuff
    \right)
    \frac{3 r^2dr}{N r_0^3} 
  },
\end{equation}
where we explicitly integrate over the radial distribution of dipoles,
and where $\expval{\cdot}$ denotes the expectation value with respect to
the direction towards the dipole. The expression readily incorporates
also an average over possibly varying directions $\vhat{n}$ of the
individual dipoles to be only briefly considered below.
By a simple substitution, we rewrite Eq.~\eqref{eq:probEz} as
\begin{equation}
  \label{eq:probEz2}
  P_{1,N}(g)
  =\frac{1}{N g^2} \, D(N g),
\end{equation}
where $D(g)$ is a geometrical factor which depends only on the
distribution of $d$:
\begin{equation}
  \label{eq:Drandom2}
  D(g)=
  \expval{
    \left| \dofstuff \right|
    \int^1_0
    \delta(u-\frac{ \dofstuff }{g}) du
  }.
\end{equation}
We observe the simple scaling of the probability distribution for
the field of a single dipole in a large volume holding on average
$N$ dipoles: $P_{1,N}(g) = N P_{1,1}(Ng)$.

\begin{figure}[htbp]
  \centering
  \includegraphics[width=8.5cm]{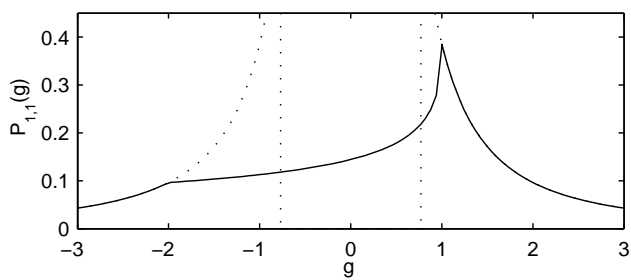}
  \caption{%
    The probability distribution $P\topp{p}_{1,1}(g)=g^{-2}
    D\topp{p}(g)$ for the field contribution from a single dipole
    parallel to the $z$-axis. The dotted curve is based on the step
    approximation $D\topp{\theta}(g)$, with the same asymptotic
    values as $D\topp{p}(g)$.  
  }
  \label{fig:dipoldist2}
\end{figure}

For dipoles parallel to the $z$-axis, $\dofstuff$ attains the
value
\begin{equation}
  \label{eq:ddpar2}
  d\topp{p}(\theta) = 1-3 \cos^2 \theta,
\end{equation}
from which we find by integration over solid angles that
\begin{equation}
  \label{eq:dpar2}
  D\topp{p}(g)=
  \frac{1}{3 \sqrt{3}}
  \begin{cases}
    2-(2+g)\sqrt{1-g}
    &    \text{if $-2<g<1$,}\\
    2 &\text{otherwise.}
  \end{cases}
\end{equation}
The fact that $D\topp{p}(g)$ assumes a constant value of
$D\topp{p}_\infty=2/3 \sqrt{3} \approx 0.3849$ for $|g|>2$ follows
from \eqref{eq:Drandom2} because $|d|$ is bounded by $2$ and it
provides $P_{1,N}$ with \concept{algebraic tails} proportional to
$g^{-2}$. This is illustrated in Fig.~\ref{fig:dipoldist2}, where
$\pcan\topp{p}$ is compared to the distribution corresponding to a
step approximation of $D\topp{p}$ with the same limiting value:
\begin{equation}
  \label{eq:dthetadef2}
  D\topp{\theta}(g)=
  \begin{cases}
    D_\infty \qquad & \text{for $|g|>2 D_\infty$}\\
    0 & \text{otherwise,}
  \end{cases}
\end{equation}
where the position of the edge is determined by the normalization of
$\pcan(g)$.

Due to the $g^{-2}$ algebraic tails, the distribution
$\pcan$ has a divergent variance and an ill-defined mean value.
This type of problems is addressed by generalized (Levy) statistics,
see e.g.~\cite{Feller66v2}, and the form of the bulk distribution
$P_\infty \equiv \lim_{N \to \infty} P_N$ can be calculated by the
generalized central limit theorem, see e.g.~chapter 17 of
\cite{Fristedt96}. We will, however, calculate $P_\infty$ directly as
the limit of $P_N$ to establish a formalism where the effect of an
excluded volume can also be obtained.

The simple scaling relation between $P_{1,N}(g)$ and $\pcan(g)$ allows
us to express $P_N$ in terms of $\pcan$ by rewriting the convolution
Eq.~\eqref{eq:Pgn2} in Fourier space as
\begin{equation}
  \label{eq:pgnFour2}
  P_N(g)=\int e^{i k g} \left( \pf_{1,1}(\tfrac{k}{N}) \right)^N \frac{dk}{2 \pi},
\end{equation}
where $\tilde{\cdot}$ denotes the Fourier transform:
$\tilde{f}(k)=\int e^{-i k g} f(g) dg$.
Eq.~\eqref{eq:pgnFour2} implies that to determine $P_N$ in the
limit of $N \to \infty$ we must know the dependence of
$\pfcan(k)$ for small $k$. We first rewrite $\pfcan(k)$ as
\begin{equation}
  \label{eq:pf_1k-=-int}
  \begin{split}
  \pfcan(k)
  = &\int e^{-i k g} g^{-2} D\topp{\theta}(g) dg\\
  &+\int e^{-i k g} g^{-2} \left(D(g)- D\topp{\theta}(g)\right) dg,
  \end{split}
\end{equation}
where the first term is conveniently rewritten as
\begin{equation}
  \label{eq:ptheta}
  1-2 |k| D_\infty \left(
    \frac{\pi}{2}-
    \int_0^{2 |k| D_\infty} \frac{1-\cos(t)}{t^2} dt
  \right).
\end{equation}
In a small $k$ expansion of the second integral of \eqref{eq:pf_1k-=-int}, 
the $0^\text{th}$ order term vanishes since $g^{-2}D(g)$ and
$g^{-2}D\topp{\theta}(g)$ are both normalized.  $D$ and $D\topp{\theta}$
are equal for all $|g|>2$ and since $D\topp{\theta}$ is even, the
$1^\text{st}$ order term yields $-i k g_c$ with $g_c$ defined as
\begin{equation}
  \label{eq:i-k-int}
  g_c= \int_{-g_0}^{g_0} g \pcan(g) dg,
\end{equation}
for any $g_0 > 2$. Collecting the two parts we find that
$\pfcan(k)=1-\pi D_\infty |k|- i k g_c+ \bigo(k^2)$.
Insertion of the expression for $D\topp{p}(g)$ leads to the value
\begin{equation}
  \label{eq:g_c=-frac29-left}
  g_c\topp{p}=
  \frac{2}{9}
  \left(
    3+\sqrt{3}\log \frac{\sqrt{3}-1}{\sqrt{3}+1}
  \right) \approx0.1598.
\end{equation}

To calculate the limit of $P_N$ for $N \to \infty$, we rewrite
\eqref{eq:pgnFour2} as $\log(\pf_N(k))= N \log \pfcan(k/N)$.
Since $\pfcan(0)=1$ and $\log(1+u)=u+\bigo(u^2)$, the leading terms of
the series expansion of $(\pfcan(k)-1)$ will dominate in the limit of $N
\to \infty$, so that
\begin{equation}
  \label{eq:rhodef}
  \log(\pf_\infty(k)) =
  - \pi D_\infty |k| - i g_c k,
\end{equation}
from which the limiting distribution follows directly:
\begin{equation}
  \label{eq:pgn2}
  P_\infty(g)=\frac{1}{\pi} \frac{\gwidth}{\gwidth^2 + (g-g_c)^2}.
\end{equation}
This Lorentzian with a half width of $\gwidth=\pi D_\infty$ and a
displacement of $g_c$ is in excellent agreement with our numerical
simulations shown in Fig.~\ref{fig:pardip2}.
The half width 
$\pi D\topp{p}_\infty \eznot \approx 5.065\, C \rho$
and central value
$g\topp{p}_c \eznot \approx 0.6692\, C \rho$ 
of the field distribution $P(\ez)$ are both proportional to the dipole
density $\rho$, and their ratio is independent of $\rho$.

The shift of the most probable value with respect to zero is
surprising when one considers the vanishing mean contribution from any
spherical shell around the origin, but it is less surprising when one
observes the probability distribution for the single dipole
contribution, shown in Figure 2.
This distribution is indeed suggestive of a shift, but its
mean is ill-defined, and \eqref{eq:i-k-int} provides the proper
procedure to obtain $g_c$ from $\pcan(g)$.

For completeness we note that, in the case of randomly oriented
dipoles, the factor $\dofstuff$ is given by
\begin{equation}
  \label{eq:drandom2}
  d\topp{r} =
    \sin \theta_1 \sin \theta_2 \sin \phi
    -2 \cos \theta_1 \cos \theta_2,
\end{equation}
where $\theta_1$ is the direction of $\vhat{r}$, $\theta_2$ is the
angle between $\vhat{r}$ and $\vhat{n}$, and $\phi$ represents the
rotations of $\vhat{n}$ around $\vhat{r}$.
Integration over these angles with the appropriate probability measure
$\sin(\theta_1) \sin(\theta_2) d\theta_1 d\theta_2 d\phi/ 8 \pi$
yields an even function $D\topp{r}(g)$ with the asymptotic limit
$D\topp{r}_\infty=\frac{1}{4}+\frac{\sqrt{3}}{24}\sinh^{-1}(\sqrt{3})
\approx 0.3450$, implying a Lorentzian distribution with a half width
$\Gamma = \pi D\topp{r}_\infty\simeq 1.083$ centered at zero field.
This is in agreement with work by Stoneham
\cite{stoneham69:shapes_inhom_broad_reson_lines}, who considered a
variety of line broadening mechanisms in solids, and identified a
Lorentzian line as the result of interaction of a single molecule with
dislocation dipoles.
More recently \cite{barkai00:levy_distr_singl_molec_line} Lorentzian
line shapes were measured for molecules embedded in low temperature
glass with a low-density distribution of dynamical defects. These
results were interpreted in terms of Levy stable distributions.

\begin{figure}
  \centering
  \includegraphics{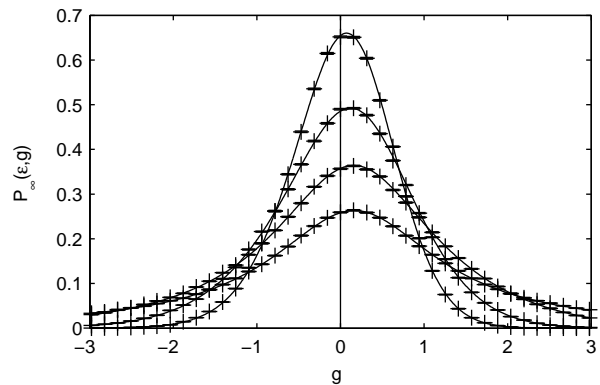}
  \caption{
    Distribution of $g$ for the case of dipoles parallel to the
    $z$-axis when an excluded volume of size $\per/\rho$ is introduced.
    Solid lines show the distribution $P_\infty(\per,g)$ given by
    \eqref{eq:rhoexform}, for $\per=0$, $0.4$, $1$, and $2$ in order
    of increasing maximum densities. Data markers are the result of
    numerical simulations.
    The distributions have vanishing mean for all values of $\per>0$,
    yet they approach the shifted Lorentzian, corresponding to $\per=0$.
  }
  \label{fig:pardipex}
\end{figure}

If the algebraic tails of $\pcan$ are truncated by some mechanism, the
distributions have finite variance, and our naive estimates of mean
values will be valid due to the central limit theorem. To investigate
whether such a truncation entirely removes the more spectacular effect
identified above, we shall compute the field distribution in the case
where we will not allow any dipoles inside an \concept{excluded
  volume} in the form of a sphere of volume $\per/\rho$ centered at
the origin. Note that $\per$ is the average number of dipoles that
would have been found in the excluded volume.
The symbols in Fig.~\ref{fig:pardipex} show the results of simulations
performed with dipoles put uniformly at random around the origin but
outside such excluded volumes, and as we reduce the excluded volume we
observe that the probability distributions converge towards the
shifted Lorentzian.
The generalized central limit theorem which applies for $\per=0$ and
$\per \to \infty$ deals with the convergence of the distribution
function for a sum of more and more random variables which all have
the {\it same} individual distribution after a suitable
rescaling. Such rescaling is not possible for intermediate values of
$\per$, which thus require a direct calculation of $P_\infty(\per,g)$.

We consider the field contribution from a single dipole placed at
random in a spherical shell with outer radius $(N+\per)^{1/3} r_0$ and
inner radius $\per^{1/3} r_0$.
Parametrizing the radius by $x=(r/r_0)^3$, the mean number of atoms
populating the sphere with radius $r$, we have by Bayes rule and the
additivity of probabilities of disjoint events that $P(\per<x<N+\per)
P(g|\per<x<N+\per)= P(x<N+\per) P(g|x<N+\per)- P(x<\per) P(g|x<\per)$,
where the first factors are simply the probabilities that a single
particle is found in the specified regions of space, and, e.g.,
$P(x<(N+\per))/P(x<\per)=(N+\per)/\per$.
Taking the Fourier transform with respect to $g$ and noting that
$\pf(k|x<x_i)=\pfcan(k/x_i)$, we obtain the following relation between
the Fourier transformed probabilities
\begin{equation*}
  \label{eq:pfnex}
  N\,\pf(k|\per<x<N+\per)=
  (N+\per)\pfcan(\tfrac{k}{N+\per})-\per\pfcan(\tfrac{k}{\per}).
\end{equation*}
We are interested in the probability $P_N(\per,g)$ that the
contributions from $N$ dipoles, all having $\per<x<N+\per$, add up to
the value $g$. Performing the convolution in
Fourier space we find that $\log \pf_N(\per,k)=N \log
\pf(k|\per<x<N+\per)$, and for $N \to \infty$ we have
\begin{equation}
  \label{eq:rhoexform}
  \log \pf_\infty(\per,k)=\log \pf_\infty(k) - \per
   (\pfcan(\tfrac{k}{\per}) -1).
\end{equation}
As shown by Fig.~\ref{fig:pardipex} this expression, which can be
evaluated numerically, is in excellent agreement with numerical
simulations.

$|\pfcan|<1$, and, by \eqref{eq:rhodef}, the term $\log \pf_\infty(k)$
will dominate Eq.~\eqref{eq:rhoexform} for $k>\per$, in agreement with
our expectation that $P_\infty(\per,g)$ should approach $P_\infty(g)$
for $\per \to 0$.
To consider the limit of $\per \to \infty$ we continue the series
expansion of $\pfcan$ to find $\log \pf\topp{p}_\infty(\per,k)=-2/5\,
\per^{-1} k^2-4 i/105\, \per^{-2} k^3 + \per \bigo((k/\per)^4)$.
Since the leading term of this expansion will dominate for $\per \gg
1$, we conclude that $P_\infty(\per,g)$ asymptotically approaches a
Gaussian distribution with variance $\text{Var}(g)=4/5\, \per^{-1}$.

In summary, we have identified a shifted Lorentzian distribution
as the probability distribution for the total field inside a
random distribution of dipoles, and we have identified a family of
distributions for the case where dipoles are not permitted inside
an excluded volume around the origin. These distributions have
vanishing mean, and they converge to Gaussian distributions in the
limit of large excluded volumes and towards the Lorentzian in the
case of small excluded volumes. It is not an inconsistency of our
results that the shifted Lorentzian is approached by distributions
with vanishing mean: a Lorentzian can be ascribed any mean value
depending on how the upper and lower limits are taken in the
integral over the distribution. There is in fact reason to
emphasize that the common procedure of fitting a spectrum to a
Lorentzian may be quite misleading if one tries to interpret a
frequency shift as the mean value of a possible perturbation of the
energy of the system.

The emergence of a non-vanishing field as the most likely result of
the surroundings of any given dipole, could have consequences
for material properties. The properties of a conventional ferromagnet
are controlled by an interplay of Coulomb forces and the Pauli
exclusion principle for electrons, which may be conveniently
represented by a spin-spin interaction term, but in novel materials,
such as recently produced carbon-nanofoams \cite{physnews04:678_1},
the actual interaction between separated magnetic dipoles may be an
important ingredient in the understanding of their collective
properties.

Heteronuclear molecules with permanent electric dipole moments have
been trapped \cite{bethlem00:elect_trapp_ammon_molec} and experiments
are planned with atomic species with particularly high magnetic dipole
moments
\nocite{goral00:bose_einst_conden_with_magnet}
\nocite{martikainen01:commen_bose_einst_conden_with}
\nocite{giovanazzi02:tunin_dipol_inter_quant_gases}
\cite{goral00:bose_einst_conden_with_magnet,martikainen01:commen_bose_einst_conden_with,giovanazzi02:tunin_dipol_inter_quant_gases}
to study polar degenerate gases and new kinds of order and collective
dynamics
%
\nocite{yi00:trapp_atomic_conden_with_anisot}
\nocite{santos00:bose_einst_conden_trapp_dipol}
\nocite{baranov02:super_pairin_a_polar_dipol}
\nocite{damski03:creat_a_dipol_super_optic}
\cite{yi00:trapp_atomic_conden_with_anisot,santos00:bose_einst_conden_trapp_dipol,baranov02:super_pairin_a_polar_dipol,damski03:creat_a_dipol_super_optic}.
Mean-field approaches in dipolar degenerate quantum gases may of
course be questionable if the mean field itself is not well defined.
Our work suggest that a critical examination of this issue is
necessary.

Highly excited Rydberg atoms in electric and magnetic fields interact
strongly \cite{fioretti99:long_range_forces_between_cold}, and fast
quantum computing \cite{jaksch00:fast_quant_gates_neutr_atoms} and
single photon generating devices
\cite{lukin01:dipol_block_quant_infor_proces} have been suggested
based on the energy shifts in atoms caused by the excitation of nearby
atoms.
Rare-earth ions in crystals have excited states with permanent
electric dipole moments, and proposals exist for quantum computing
within such a system which are also based on large
\cite{ohlssona02:quant_comput_hardw_based_rare} or small
\cite{longdell03:selec_ensem_rare_earth_quant} shifts in absorption
frequency of target ions caused by excitation of a nearby control-ion.
In the rare-earth system Lorentzian broadening of spectrally hole
burnt structures has been observed when ions at different frequencies
are excited \cite{nilsson02:initial_exper_concer_quant_infor}, and we
imagine that this can be an ideal system to study the broadening and
the shift systematically, as the density of perturbing dipoles can be
varied by the exciting laser system.

Extension of the analysis, e.g., to time-dependent fields and to
higher order multipole fields seems very interesting.
Von Neuman and Chandrasekhar considered the fluctuating gravitational
forces in a stellar medium, see
\cite{chandrasekhar43:stoch_probl_physic_astron}. As a curiosity we
note that the time derivative of these forces at any given time
behaves like a sum of dipole fields, and hence a massive object moving
through a static random mass distribution may experience a force with
a time derivative given by the shifted Lorentzian.

\begin{acknowledgments}
  The authors wish to thank Francois Bardou for stimulating
  discussions and for useful references.
  This work was funded by the European Union IST-FET programme
  ESQUIRE.
\end{acknowledgments}


\end{document}